\documentclass[
prb,
aps,
onecolumn,
twocolumn,
amsmath,
amssymb,
superscriptaddress,
floatfix
]{revtex4-2}
\usepackage[T1]{fontenc}
\usepackage[english]{babel}
\usepackage{graphicx}
\usepackage{dcolumn}
\usepackage{amsmath}
\usepackage{amssymb}
\usepackage{bm}
\usepackage{multirow}
\usepackage[normalem]{ulem}
\usepackage[usenames,dvipsnames,svgnames,table]{xcolor}
\usepackage{braket}
\usepackage{subfigure}
\usepackage[colorlinks]{hyperref}
\usepackage{physics}

\newcommand{\refeqn}[1]{(\ref{#1})} 
\newcommand{\ts}{\:\!} 
\newcommand{\Hamil}{\mathcal{H}} 
\newcommand{\vect}[1]{\textbf #1}

\usepackage{accents}
\makeatletter
\DeclareRobustCommand{\cev}[1]{%
  \mathpalette\do@cev{#1}%
}
\newcommand{\do@cev}[2]{%
  \fix@cev{#1}{+}%
  \reflectbox{$\m@th#1\vec{\reflectbox{$\fix@cev{#1}{-}\m@th#1#2\fix@cev{#1}{+}$}}$}%
  \fix@cev{#1}{-}%
}
\newcommand{\fix@cev}[2]{%
  \ifx#1\displaystyle
    \mkern#23mu
  \else
    \ifx#1\textstyle
      \mkern#23mu
    \else
      \ifx#1\scriptstyle
        \mkern#22mu
      \else
        \mkern#22mu
      \fi
    \fi
  \fi
}
\makeatother

\begin{document}

\title{Temperature dependent energy gap for Yu-Shiba-Rusinov states at the quantum phase transition}
\author{Andreas Theiler}
 \affiliation{Department of Physics and Astronomy, Uppsala University, Box 516, 
751 20 Uppsala, Sweden}

\author{Christian R. Ast}
\affiliation{Max-Planck-Institut f\"ur Festk\"orperforschung, Heisenbergstra\ss e 1, 705 69 Stuttgart, Germany}

\author{Annica M. Black-Schaffer}
\affiliation{Department of Physics and Astronomy, Uppsala University, Box 516, 
751 20 Uppsala, Sweden}

\date{\today}

\begin{abstract}
    Motivated by recent experiments, which allow for fine tuning of the effective magnetic interaction between the impurity and the superconductor, we investigate the regime around the quantum phase transition where the system's ground state changes from a weakly coupled free spin to a screened spin regime. At this transition we find that the Yu-Shiba-Rusinov (YSR) states remain at finite energies at low temperatures, thereby generating a gap in the spectrum, which is inconsistent with  predictions of the original YSR theory. We investigate various gap-generating scenarios and determine that the local suppression of the order parameter, only captured by self-consistent calculations, generates the gap.
\end{abstract}

\maketitle

\section{Introduction}
\label{sec:introduction}

Superconductivity and magnetism are generally considered antagonistic competing orders in condensed matter.
In particular, if global magnetic interactions are introduced into a conventional superconductor, superconductivity will be suppressed and eventually vanishes as the magnetic coupling is increased. 
However, if the magnetic interaction is introduced locally by a single, isolated magnetic impurity, a new type of sub-gap impurity state is introduced into the system. 
These Bogoliubov quasiparticle states were first predicted by Yu, Shiba and Rusinov \cite{luh1965bound, shiba1968classical, rusinov1969superconductivity, rusinov_theory_1969} and are commonly known as YSR states. Due to the particle-hole symmetry of the Bogoliubov spectrum, the YSR states always appear symmetrically around zero energy.

The sub-gap nature of the YSR states serves as an intriguing method for investigating and manipulating local characteristics of superconductivity. For example, YSR states have been suggested as potential components for creating topological superconducting heterostructures that support Majorana bound states at the end points of adatomic chains of magnetically coupled impurities \cite{nadj-perge_observation_2014, ruby_end_2015, kezilebieke_coupled_2018, ruby_wave-function_2018} or through the interaction of two YSR states within quantum dots \cite{zatelli_robust_2024}.
Furthermore, their application as quantum bits, or qubits, is discussed \cite{mishra_yu-shiba-rusinov_2021, pavesic_qubit_2022, steffensen_ysr_2024}.

Intriguingly, YSR states also exhibit a quantum phase transition (QPT) when tuning the interaction between the spin of the magnetic impurity and the superconducting substrate. When this interaction is below a critical strength, the system is in a free spin regime and the ground state is a spin doublet (assuming the impurity spin is $\frac{1}{2}$), and thus the superconducting condensate only couples weakly to the magnetic impurity.
In contrast, above the critical interaction strength of the QPT, the impurity spin is screened by breaking a Cooper pair in the condensate, resulting in a spin singlet ground state.
The QPT is thus marked by the transition of the ground state from a spin doublet to singlet, while at the same time changing the influence on the superconducting condensate. In terms of the energy spectrum, the QPT is marked by the YSR states crossing zero energy, thereby interchanging the occupation of the YSR states.

Experimentally, YSR states are routinely observed in scanning tunneling microscopy (STM) \cite{heinrich_single_2018}.
However, the QPT can be experimentally difficult to observe, as the magnetic interaction is set by the choice of magnetic impurity and superconductor and thus difficult to manipulate in a given material combination. 

In fact, it is very difficult to determine from STM spectra, on which side of the QPT the YSR state resides, so that this could only be done indirectly \cite{farinacci_tuning_2018,huang_quantum_2020}.
An innovative experimental development that allows for fin-tuning the magnetic interaction is to exploit the atomic forces between tip and sample. By changing the tip-sample distance, the atomic forces push or pull on the impurity thereby changing its coupling to the superconductor. This effect has been exploited in a number of different systems and allows for careful investigation of the QPT \cite{ternes2006scanning, farinacci_tuning_2018, brand2018electron, malavolti_tunable_2018, kezilebieke_observation_2019, huang_quantum_2020, karan_superconducting_2022, huang_universal_2023}.

In this work we capitalize on the recent experimental advancement on tuning YSR states and investigate in detail the behavior of the YSR states, at and around the QPT. We are particularly focusing on the possibility and physical reason for a finite energy gap at the QPT, such that the YSR states never reach zero energy.
Such a gap in the energy spectrum at the QPT has been indicated in several earlier numerical works, but not much further elaborated upon.
To the best of our knowledge, the first instances where a gap has been noted in the energy spectrum is Ref.~\onlinecite{salkola1997spectral,flatte1997localDefects, flatte1997localMagnetic}, which modeled a magnetic impurity in the classical spin limit and relaxed the local order parameter of the superconducting substrate.
A putative finite gap can also be found when closely investigating the results of later works using similar methods \cite{ meng2015superconducting, pershoguba2015currents}.
In another model comprising two coupled impurity spins, the presence of a gap has also been observed for specific parameter choices, but this effect does not account for an energy gap for a single impurity spin \cite{uldemolins_hunds_2024}.
Experimentally, a finite energy gap has also been reported in Ref.~\cite{karan_superconducting_2022} for a single magnetic impurity, but, again, the underlying mechanism was not substantially elaborated upon, as the gap was not the focus of that work.
Here we extend on these earlier results on a finite energy gap for the YSR states and show both how and why a gap can appear in the YSR spectrum, using similar modeling to the afore-mentioned numerical works.

The remainder of this work is organized as follows.
In Section \ref{sec:YSR}, we motivate and explain the models we use to describe a single magnetic impurity and its YSR states.
We then formulate two general and exhaustive routes to achieve finite-energy YSR states at the QPT. 
The first route, described in Section \ref{sec:YSRcoupling}, is based on the possibility of adding interactions to the model that may open a gap.
We demonstrate that even with such additional model terms, the YSR states still cross at zero energy, with no gap-opening possible due to symmetry restrictions set by the superconducting pairing.
In Section \ref{sec:gapselfcons}, we explore the second possible route, which consists of forcing a discontinuous change of the YSR spectrum at the QPT. 
Such discontinuous change can only be possible in the superconducting order parameter, since all other parameters are either constant or changes smoothly across the QPT.
In fact, it is well known that the superconducting order parameter suffers a local suppression at and very close to magnetic impurities \cite{rusinov1969superconductivity,heinrichs1968spatial, schlottmann1976spatial, kim1993spatial, balatsky_impurity-induced_2006}, even to the degree that the order parameter can accumulate a $\pi$-shift at the magnetic impurity past the QPT \cite{salkola1997spectral, flatte1997localDefects, flatte1997localMagnetic, balatsky_impurity-induced_2006, meng2015superconducting, pershoguba2015currents, graham2017imaging, bjornson2017superconducting, theiler2019majorana}.
This local suppression sets in abruptly at the QPT and we show that due to the self-consistent relationship between the superconducting order parameter and the energy spectrum, this order parameter behavior can result in a finite YSR state energy at the QPT.
We further find that the YSR energy gap occurs only at low temperatures and is washed out by thermal occupation of the positive energy YSR states at higher temperatures.
We also highlight how the magnetization of the system can be used to track this gap-opening phenomena.
Finally, in Section \ref{sec:conclusions} we summarize our results.

\section{Modeling YSR states}
\label{sec:YSR}
\begin{figure}%
\centering
\includegraphics[width=0.475\textwidth]{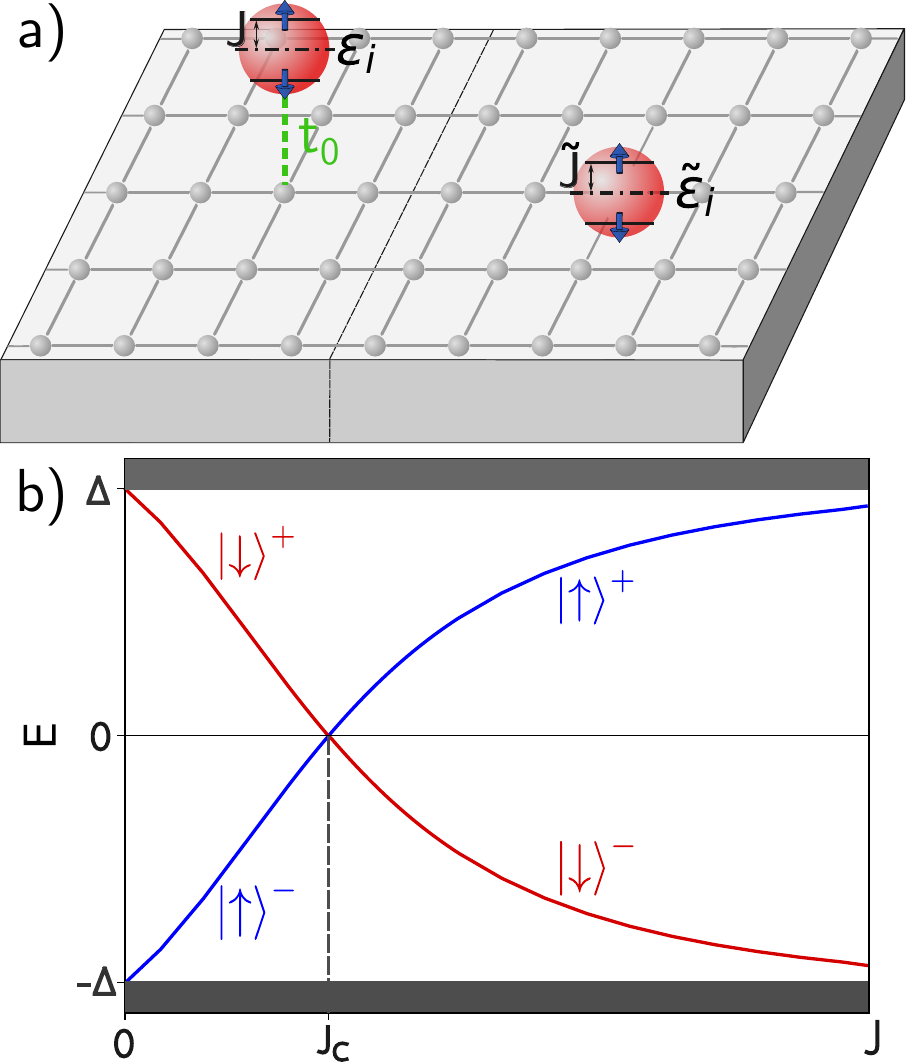}%
\caption{a) Schematic picture of a superconducting substrate with indicated square lattice and with an absorbed magnetic impurity (red dot), treated within (left) the effective Anderson impurity model, with impurity on-site energy $\varepsilon_{i}$ and spin exchange coupling $J$ for the impurity, and with the impurity coupling to the substrate through the hopping amplitude $t_0$ and (right) classical Kondo model with effective spin exchange coupling $\tilde{J}$ and additional onsite energy of $\tilde{\varepsilon}_{i}$, both induced directly in the substrate.
b) Ideal YSR state energies $\epsilon_{YSR}$ with states labeled according to their spin-polarization relative to the impurity spin, where superscript indicates occupied ($-$) and unoccupied ($+$) at zero temperature. The critical coupling $J_C$ marks the QPT.}%
\label{fig:toy_model}%
\end{figure}

We consider a single magnetic impurity deposited on the surface of a conventional spin-singlet $s$-wave superconducting substrate.
To characterize such an impurity, two models are frequently employed: an Anderson impurity model in the mean-field limit and the Kondo model in the classical spin limit, see Fig.~\ref{fig:toy_model}a).
The Anderson impurity model models an impurity as a quantum level system with a repulsive Hubbard interaction.
Knowing that the impurity is already magnetic due to the Hubbard interaction, we can describe it within an effective mean-field picture in terms of an magnetic mean field $J$ in the impurity Hamiltonian $\Hamil_{i}$.
Moreover, we add a possible onsite energy $\varepsilon_i$ of the impurity level to $\Hamil_{i}$.
However, note that both parameters $J$ and $\varepsilon_i$ are not explicitly calculated from the original Anderson impurity model, but we treat them as free parameters, which allows us to explore all possibilities for the interaction with the superconducting substrate. 
We couple the impurity to the superconducting substrate, described by $\Hamil_{sc}$, through a coupling Hamiltonian $\Hamil_{c}$ with the hopping element $t_0$.

The Anderson impurity model, within the mean-field treatment described above, from here on referred to as the Anderson impurity model, can be simplified further in the classical spin limit by modeling the effective interaction between the impurity and the substrate as a local magnetic scattering amplitude $\tilde{J}$ directly present at a single site of the substrate.
This results in the addition of an additional impurity term to the Hamiltonian substrate $\Hamil_{sc}$, eliminating the need to explicitly consider the impurity-substrate coupling.
Overall, this model description is the Kondo impurity model in the classical spin limit for the magnetic impurity.

In general, both impurity models described above, quantitatively describe the same physics, also highlighted by the fact that they are directly linked via a Schrieffer-Wolff transformation \cite{schrieffer1966relation} in the strong coupling limit.
This transformation combines the effects of the coupling between the impurity and the substrate with the energy separation of the impurity levels, into the single scattering amplitude $\tilde{J}$.
In this work we consider the mean-field Anderson impurity model when discussing general properties of YSR states and possibilities for an energy gap in the following Section~\ref{sec:YSRcoupling}, but then we need to revert to the slightly simpler Kondo limit for the numerically demanding self-consistent calculations in Section~\ref{sec:gapselfcons}. 
Overall, we find our results and conclusions to be fully consistent between the two models, as also guaranteed by their direct link \cite{schrieffer1966relation}.
This involves numerical verification of key results between the two models.

Before proceeding we note that the impurity treatment in this work, as in many others, do not incorporate possible quantum impurity behavior, e.g. see review \cite{balatsky_impurity-induced_2006} and references therein or for a comparison between different models \cite{martin-rodero_josephson_2011, martin-rodero_andreev_2012, zitko_quantum_2018}.
Such a quantum impurity requires a many-body treatment, which demands computationally very expensive numerical methods, such as numerical renormalization group (NRG) \cite{wilson1975NRG} or quantum Monte Carlo methods \cite{RevModPhys.83.349, PhysRevB.81.024509, PhysRevLett.108.227001}. Still, the single-particle impurity models, or classical spin models, we use in our work been shown to reproduce important features of the full quantum impurity problem \cite{martin-rodero_josephson_2011, martin-rodero_andreev_2012, zitko_shiba_2015, kadlecova2017QuantumDot, zitko_quantum_2018, kadlecova2019PracticalGuide}.
Most importantly, classical spin models are well motivated  both for magnetic impurities with high spin or for systems with axial magnetic anisotropy \cite{zitko_quantum_2018}, and has also been shown to produce results in good agreement with recent experimental findings also in spin-1/2 systems \cite{huang_quantum_2020, karan_superconducting_2022}.
Finally, since our main results are based on a careful self-consistent treatment of the local order parameter, which also requires high resolution with respect to the magnetic coupling around the QPT, we are computationally limited to consider the classical spin limit.
However, we note that an extension to quantum impurities is in principle feasible within an NRG treatment \cite{satori_numerical_1992}, albeit very computationally expensive.

We next provide more specific details for the models described above before solving for the energies of the YSR states.
Following earlier work \cite{villas_interplay_2020, huang_quantum_2020, karan_superconducting_2022}, we start with the mean-field Anderson impurity description, where the Hamiltonian for the magnetic impurity $\Hamil_{i}$ can be described, using the Nambu spinor $\Psi_i^\dagger = \left( d_\uparrow^\dagger, d_\downarrow, d_\downarrow^\dagger, -d_\uparrow \right)$, as 
\begin{equation}
    \label{eqn:impurity_level}
    \begin{split}
           \Hamil_{i} &= \Psi_i^\dagger h_{i} \Psi_i \\
           \text{with} \quad h_{i} &=  \varepsilon_{i} \sigma_0 \otimes \tau_3 + J \sigma_3 \otimes \tau_0.
    \end{split}
\end{equation}
Here $\varepsilon_{i}$ describes the effective on-site energy on the impurity site and $J$ the mean-field exchange coupling responsible for the magnetic interaction and thereby energy splitting between the two spin states of the impurity.
We choose the direction of the exchange coupling to be along the $\hat{z}$ direction, without loss of generality.
Further, the Pauli matrices $\sigma$ and $\tau$ act in the spin and electron-hole (Nambu) spaces, respectively.
The impurity level is absorbed on a conventional spin-singlet $s$-wave superconducting substrate described by the Bogolioubov-de Gennes (BdG) Hamiltonian
\begin{equation}
\begin{split}
        \Hamil_{sc} &= \sum_{\vect{k}} \Psi_{s, \vect{k}}^\dagger h_{sc, \vect{k}} \Psi_{s,\vect{k}} \\
        \text{with} \quad h_{sc, \vect{k}} &= \varepsilon_{\vect{k}} \sigma_0 \otimes \tau_3 + \Delta \sigma_0 \otimes \tau_{1},
\end{split}
\end{equation}
where the Nambu spinor of the substrate is given by $\Psi_{s,\vect{k}} ^\dagger = \left( c_{\vect{k},\uparrow}^\dagger, c_{-\vect{k},\downarrow}, c_{\vect{k},\downarrow}^\dagger, -c_{-\vect{k},\uparrow} \right)$ and define the basis for the corresponding, momentum $\vect{k}$ dependent, substrate Hamiltonian $h_{sc, \vect{k}}$ of the substrate. 
Here, $\varepsilon_{\vect{k}}$ is the kinetic energy and $\Delta$ is the superconducting gap parameter of the substrate.
As we assume spin-singlet $s$-wave superconductivity, there is no $\vect{k}$-dependence for $\Delta$. Later we allow $\Delta$ to develop a spatial dependence as a reaction to the magnetic impurity.
Assuming a perfect crystal and the impurity being fully localized in space, the coupling between the substrate and the impurity is equal for all momenta in the substrate such that the coupling Hamiltonian is written as
\begin{equation}
    \label{eqn:coupling_hamiltonian_analytical}
    \begin{split}
        \Hamil_{c} &= \sum_{\vect{k}} \left( \Psi_i^\dagger V_{is} \Psi_{s, \vect{k}} + \Psi_{s, \vect{k}}^\dagger V_{si} \Psi_i \right),\\
        \text{with} \quad V_{is} &= -t_0 \sigma_0 \otimes \tau_3,
    \end{split}
\end{equation}
where $V_{is}$ is the coupling between the substrate with $V_{is} = V_{si}^\dagger$.
Here, $t_0$ encodes the strength of the coupling, or equivalently hopping amplitude, between the impurity and the substrate and is assumed to have no spin dependence. The total Hamiltonian for the mean-field Anderson impurity model of an impurity on a superconducting substrate therefore reads: $\Hamil = \Hamil_i + \Hamil_{sc} + \Hamil_c$.

\subsection{YSR energies}
The Hamiltonian of the mean-field Anderson impurity model on a superconductor $\Hamil$ can in principle be solved numerically by exact diagonalization for all eigenstates and energies and then the YSR states can be isolated as the only in-gap states.
To instead analytically extract the YSR  states we employ the Green's function method and calculate the local dressed (retarded) Green's function of the impurity using the Dyson equation
\begin{equation}
    \label{eqn:dyson_equation}
    G_{ii}\left(E\right) = g_{ii} + g_{ii} \Sigma  G_{ii}.
\end{equation}
Here the undressed local impurity Green's function is given by $g_{ii} = \left( E  - h_i \right)^{-1}$, while the self energy $\Sigma\left(E\right) = V_{is} g_{ss} V_{si}$ involves the coupling from Eq.~\refeqn{eqn:coupling_hamiltonian_analytical} and the local Green's function of the superconducting substrate $g_{ss}$.
This momentum dependent (retarded) Green's function is given by $g_{ss,\vect{k}}\left(E\right) = \left( E - h_{sc, \vect{k}} \right)^{-1}$. 
The local Green's function is then calculated by transforming the summation over $\vect{k}$ into a sum over the dispersion relationship $\varepsilon_\vect{k}$, yielding
\begin{equation}
\label{eqn:local_bare_sc_greens_fct}
    g_{ss}\left( E \right) = -\frac{N_0 \pi}{\sqrt{\Delta^2 - E ^2}} \left( E \sigma_0 \otimes \tau_0 + \Delta \sigma_0 \otimes \tau_1 \right),
\end{equation}
where $N_0$ is the normal state density of states at the Fermi level of the substrate. 
For clarity we have suppressed the addition of an infinitesimal $i\delta$ term above in the denominators of the Green's functions.

Extracting the impurity Green's function $G_{ii}(E)$ by solving Eq.~\eqref{eqn:dyson_equation} results in the YSR limit, where we assume $J \gg \Delta$, in \cite{villas_interplay_2020, huang_quantum_2020, karan_superconducting_2022}
\begin{widetext}
\begin{equation}
\begin{split}
    \label{eqn:greens_fct_classical_ysr}
            G_{ii} \left(E \right) &= \begin{pmatrix}
                G_{ii, \uparrow\uparrow} \left(E \right) & 0 \\
                0 & G_{ii, \downarrow \downarrow} \left(E \right)
            \end{pmatrix}, \\
            G_{ii, \sigma \sigma} \left(E \right) &= \frac{1}{D_\sigma(E)} \left( \Gamma E \tau_0 + \sqrt{\Delta^2 - E^2} \left[ \left(E  - J \left(\sigma_3\right)_{\sigma \sigma} \right) \tau_0 + \varepsilon_{i} \tau_3 \right] + \Delta \Gamma \tau_1 \right), \\
            D_\sigma(E) &= \Gamma E \left[E - J (\sigma_3)_{\sigma\sigma}\right] + \sqrt{\Delta^2 - E^2} \left[\left(E - J (\sigma_3)_{\sigma \sigma}\right)^2 - \varepsilon_{i}^2 - \Gamma^2 \right],
\end{split}
\end{equation}
\end{widetext}
where we define $\Gamma = N_0 \pi t_0^2$ as the coupling rate.
The energies of the YSR states are then simply given by the poles of the impurity Green's function, i.e.~$D\left(E\right) = 0$, lying within the proximitized superconducting gap, which finally yields \cite{villas_interplay_2020, huang_quantum_2020, karan_superconducting_2022}
\begin{equation}
    \label{eqn:classical_ysr_state_energy}
    \epsilon^{0, \pm}_{YSR} = \pm \Delta \frac{\Gamma^2 + \varepsilon_{i}^2 - J^2}{\sqrt{\left( \left(\varepsilon_{i} + J\right)^2 + \Gamma^2 \right) \left( \left(\varepsilon_{i} - J\right)^2 + \Gamma^2 \right)}}.
\end{equation}
Here we use the superscript $0$ to indicate that this results holds in the idealized YSR limit of $J \gg \Delta$ and for the terms used in Eqn. (\ref{eqn:impurity_level}-\ref{eqn:coupling_hamiltonian_analytical}). 

The YSR state energies $\epsilon^{0, \pm}_{YSR}$ in Eq.~\eqref{eqn:classical_ysr_state_energy} are symmetrically positioned around zero energy, as expected due to the particle-hole symmetric BdG energy spectrum of a superconductor, and by tuning $J$, it is clear that the energies can become zero.
To illustrate this explicitly, we plot the general behavior of the YSR energies in Fig.~\ref{fig:toy_model}b) as a function of $J$.
Here we label and color the two YSR states according to their spin orientations and use superscripts ($\pm$) to indicate unoccupied and occupied states, respectively.
As seen, there is a smooth and linear crossing of the YSR states at zero energy at a critical coupling $J_C$.
The behavior of the YSR energies can be understood as a direct consequence of the block structure in Eq.~\eqref{eqn:greens_fct_classical_ysr}:
The two YSR states originate from the two different blocks, where each solution varying continuously as a function of $J$.
Since the YSR energies are also symmetric around zero, this necessarily leads to a smooth crossing at zero energy occurring at a specific $J_C$ and notably no energy gap in the spectrum. 
The fact that each YSR state is generated from a separate spin block in Eq.~\eqref{eqn:greens_fct_classical_ysr} also automatically sets the spin character of the states, and thereby the spin character of the occupied YSR state is immediately flipped at $J_C$ at zero temperature. 
In fact, in the following we use as definition of $J_C$ also at finite temperatures to be the point where the spin polarization of the predominantly occupied sub-gap state changes. Due to the spin-flip of the occupied state, the crossing point $J_C$ also marks a QPT in which the ground state of the system changes:
At $J<J_C$ there is an unscreened spin at the impurity site, while beyond the QPT one Cooper pair in the bulk superconductor is broken up resulting in superconducting condensate screening the spin \cite{balatsky_impurity-induced_2006}. 

\subsection{YSR gap possibilities}
Having established the idealized properties of the YSR states, we next consider the options for opening a gap in the YSR energy spectrum, at the QPT.
Generally speaking, there are only two modification that can be introduced in the ideal model to produce a gap in the spectrum.
\begin{figure}%
\centering

\includegraphics[width=0.45\textwidth]{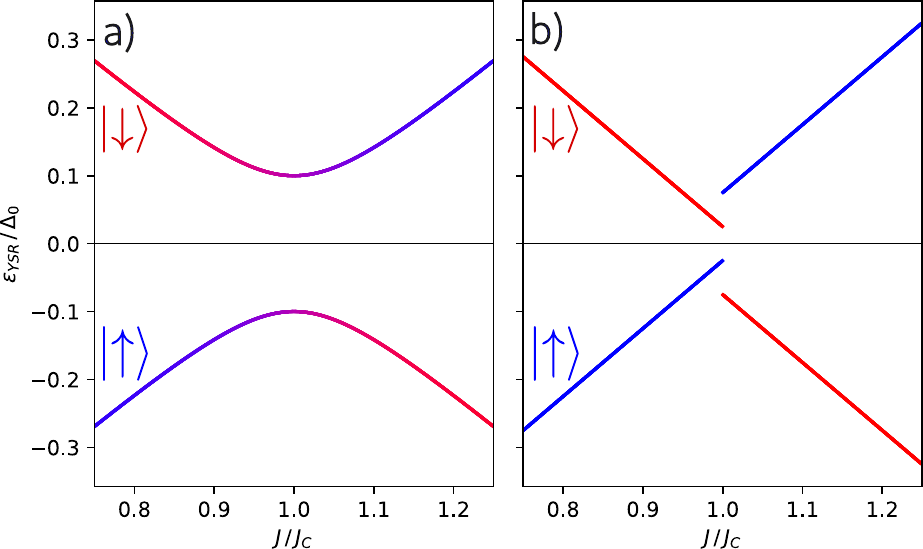}
\caption{
Schematic illustration of the two scenarios for opening a gap in the YSR energy spectrum around the QPT as a function of effective coupling $J$.
Colors indicate the two spin polarization directions of the YSR states, with spin-down (red) and spin-up (blue). 
a) Off-diagonal (spin-dependent) interaction with regard to the local impurity Green's function Eq.~\refeqn{eqn:greens_fct_classical_ysr} causing mixing of the spin polarization.
b) Indirect coupling between the two YSR states, caused by a local, discontinuous, change of each spin component of the Green's function, specifically a local suppression of the superconducting order parameter.
This does not induce mixing of the YSR spin polarizations, but the energy spectrum can still become discontinuous at $J_C$.
With the critical $J_C$ set by the switch in spin-polarization of the occupied state, it coincide with the location of the minimum gap in both scenarios.
}
\label{fig:schematic_gap_types}%
\end{figure}

The first scenario uses additional terms in the original Hamiltonian $\Hamil$ that introduce off-diagonal components, or interactions, in Eq.~\refeqn{eqn:greens_fct_classical_ysr}, thereby coupling the two diagonal blocks of the local Green's function, $G_{ii,\uparrow \uparrow}$ and $G_{ii,\downarrow \downarrow}$.
This then directly provides a coupling between the two spin orientations of the YSR states and thereby the possibility of a gap.
Such a spin-dependent interaction can thus result in a behavior similar to an avoided level crossing in any band structure, with associated mixing of the spins close to the avoided level crossing, or equivalently the QPT at $J_C$.
In Fig.~\ref{fig:schematic_gap_types}a) we schematically illustrate the effects of such an additional coupling on the YSR energy spectrum.
As an example, a putative candidate for this interaction would be some type of spin-orbit coupling as it may provide a mixing between different spin orientations.
However, common Rashba spin-orbit coupling in the substrate has been reported to not cause a gap in the spectrum \cite{pershoguba2015currents} and as such, more complicated forms are needed.
We analyze all possible interactions that may result in an energy gap in Section \ref{sec:YSRcoupling}, where we establish that there is no allowed way of generating a gap.

The second scenario to produce a gap in the spectrum is to in some indirect way couple the two diagonal blocks of the local Green's function in Eq.~\refeqn{eqn:greens_fct_classical_ysr}.
Even if there is no direct coupling present in this second scenario, it may still be possible to provide a change of the ingoing parameters that affects both blocks, such that the crossing at $J_C$ is no longer smooth, or even continuous, and, as a result, a gap may be generated. We illustrate schematically such a behavior in  Fig.~\ref{fig:schematic_gap_types}b), where there is a jump in the energy levels for both the upper and lower YSR states, such that no state actually cross zero energy. 
Obviously this scenario requires a dramatic modification to the Green's function exactly at $J_C$ and may not be what is at first expected by just continuously tuning a single parameter $J$ for a single impurity in a thermodynamically large system. However, it is already well-established that the superconducting order parameter $\Delta(\vect{x}_0)$ at the impurity site $\vect{x}_0$, is heavily suppressed and even shift non-continuously to a negative value at the QPT, denoted as a $\pi$-shift in the order parameter \cite{salkola1997spectral, flatte1997localDefects,morr_impurities_2006, balatsky_impurity-induced_2006, meng2015superconducting, pershoguba2015currents, graham2017imaging, bjornson2017superconducting, theiler2019majorana}. 
Since the order parameter in turn determines the YSR energy levels through $\Hamil$, such drastic change of the order parameter may be a candidate to also induce a discontinuous behavior of the energy levels at $J_C$, akin to Fig.~\ref{fig:schematic_gap_types}b), resulting in a gap in the spectrum.
A hint of such gap is in fact present already in early numerical data \cite{salkola1997spectral} as well as later works \cite{flatte1997localMagnetic, flatte1997localDefects, morr_impurities_2006, meng2015superconducting, pershoguba2015currents}, but has to our knowledge not been further discussed in the literature.
We finally note that in this second scenario there is no mixing between the two YSR states and the spin polarization of each state therefore stays constant.
We show in Section \ref{sec:gapselfcons} that this scenario of opening a gap can occur along with a strong temperature dependence.

\section{Gap from interactions between YSR states}
\label{sec:YSRcoupling}
From Eq.~\refeqn{eqn:greens_fct_classical_ysr} we see directly that the impurity Green's function has a diagonal block structure, with each of the two blocks encoding separately one of the spin-polarized YSR peak in the spectrum.
In order to couple the YSR states from each block at the QPT, and thereby produce a gap in the energy spectrum, additional terms that are off-diagonal in Eq.~\refeqn{eqn:greens_fct_classical_ysr} thus have to be present.
However, not all of such off-diagonal terms produce a gap in the energy spectrum, for example, if they vanish at the QPT due to their energy dependence.
Below we discuss all possible such off-diagonal terms for producing a gap, starting with terms possible to add at the impurity site, i.e.~by modifying Eq.~\refeqn{eqn:impurity_level}, and then also adding terms that modify the coupling Hamiltonian Eq.~\eqref{eqn:coupling_hamiltonian_analytical}. 
We note that we can here omit a discussion on the influence of the substrate as we are interested modeling the effects of a conventional spin-singlet $s$-wave superconductor, which has no other possible pairing terms. We further note that all other (normal-state) substrate effects, such as various spin-orbit coupling terms, can also be included in an extended treatment of the impurity-substrate coupling Hamiltonian and is hence already effectively included in our treatment. 

Starting by considering the impurity site itself, originally modeled by Eq.~\refeqn{eqn:impurity_level}, there are no additional terms possible that would introduce a meaningful coupling between the two YSR states. The reason is as follows.
Such a term needs to necessarily involve $\sigma_1$ or $\sigma_2$ in order to couple the two spins, as the other two possibilities, $\sigma_0$ and $\sigma_3$, preserve the block structure of Eq.~\refeqn{eqn:greens_fct_classical_ysr}. 
A term proportional to $\sigma_{1,2} \otimes \tau_0$ only introduces a spin rotation of the magnetic impurity and an additional energy separation of opposite spin orientations.
However, since all other terms that we have defined so far are invariant under spin rotation, such a term does not change the energy spectrum, apart from changing the effective magnetic coupling.
Furthermore, terms proportional to $\sigma_{1,2} \otimes \tau_{1,2}$ represent $s$-wave spin-triplet superconducting terms, which are forbidden pairings due to fermionic exchange rules \cite{linder_odd-frequency_2019,cayao_odd-frequency_2020,triola_role_2020}.
The last possibility for on-site terms are terms proportional to $\sigma_{1,2} \otimes \tau_3$, which cancel due to the fermion commutation rules and therefore, again, are forbidden to be included in the Hamiltonian. 

Next we consider adding extra terms to the coupling between the substrate and the impurity site by changing the coupling to
\begin{equation}
    \label{eqn:Vis}
      \hat{V}_{is} = -t_0 \sigma_0 \otimes \tau_3 + \sum_{i,j = 0}^{3} \alpha_{i\ts j} \sigma_i \otimes \tau_j,
\end{equation}
where additional contributions are introduced by the coefficients $\alpha_{i\ts j}$.
Depending on the choice of indices $i,j$, these contributions can represent spin-orbit coupling effectively introduced on the bond to the impurity, spin selective hopping to the impurity, or an unconventional superconducting interaction introduced by the impurity.
For these possible additional interactions, we again conclude that only interactions proportional to $\sigma_{1,2}$ are possible candidates to create a gap in the YSR spectrum.
Furthermore we note that, due to the doubling of terms in the BdG Hamiltonian, some terms in the coupling Hamiltonian Eq.~\refeqn{eqn:Vis} cancel after the summation over $\vect{k}$ present in Eq.~\eqref{eqn:coupling_hamiltonian_analytical}. The summation over all $\vect{k}$ is required as soon as the substrate is translationally invariant with inversion symmetry in the plane, as we expect for a clean conventional superconductor.
For example, for $\alpha_{1\ts 3}$ we generate the terms $\alpha_{1\ts 3}^* c^\dagger_{\downarrow, -\vect{k}} d_{\uparrow} + \alpha_{1\ts 3} d_{\uparrow} c^\dagger_{\downarrow, \vect{k}}$.
This term cancels when summing over $\vect{k}$ if $\alpha_{13}$ is real.
On the other hand, if $\alpha_{13}$ is imaginary, the term does not cancel, but we find that it also does not produce a gap in the spectrum.
We find a similar behavior for the remaining terms, proportional to $\sigma_{1,2}$, where the terms either cancel with the $\vect{k}$-summation or do not produce a gap.

Based on the above results, we conclude that for ideal YSR states, none of the additional terms in Eq.~\refeqn{eqn:Vis} produce a gap. However, there may still exist altered situations where the summation over $\vect{k}$ does not lead to a full cancellation as it requires inversion symmetry in the substrate plane.
This can for example be the case if the superconductor is found to be very dirty, where we cannot necessarily expect inversion symmetry to prevail, not even on an average level, or, if the coupling between the impurity and the substrate is not fully localized (i.e.~it has a $\vect{k}$ dependence) and additionally violates the inversion symmetry between $\vect{k}$ and $-\vect{k}$. In these cases the summation over $\vect{k}$ in Eq.~\eqref{eqn:coupling_hamiltonian_analytical} may not cause a full cancellation for some of the additional coupling terms in Eq.~\eqref{eqn:Vis}. 
For the sake of completeness, we report the results for creating an energy gap for such altered situations in Appendix \ref{subsec:YSR_interactions_appendix}.
We further point out that the induced magnitude of the energy gap at the QPT in these situations is generally temperature independent, unless the coefficients $\alpha_{i\ts j}$ happens to have an explicit temperature dependency.
This in sharp contrast to the gap that is caused from the indirect coupling through the self-consistently calculated local order parameter, that we discuss in the next Section \refeqn{sec:gapselfcons}.

\section{Gap from local suppression of superconductivity}
\label{sec:gapselfcons}
Having exhausted the possibilities to directly couple the two low-lying YSR states to each other and thereby opening a gap in the YSR spectrum, we next investigate the other possibility for generating a gap, presented schematically in Fig.~\ref{fig:schematic_gap_types}b). We do this by performing fully self-consistent calculations of the superconducting order parameter. The aim is to investigate if and how a gap in the spectrum may be generated by the drastic local changes of the superconducting order parameter occurring due to a magnetic impurity. We note that such calculations, which allow the superconducting substrate to respond to the magnetic perturbation, should always be performed to accurately describe a magnetic impurity. 

\subsection{Self-consistent modeling}
\label{subsec:selfconsmodel}
To be able to perform accurate numerical modeling, we, for this part, resort to the classical Kondo limit of the magnetic impurity, which reduces the numerical complexity while still capturing the same essence of the magnetic impurity.
To avoid any misunderstandings, we will in this Section use variables with a tilde, such as $\tilde{J}$, to highlight the use of the Kondo model, instead of the Anderson impurity model in the previous Section.
As such, we consider a conventional spin-singlet $s$-wave  superconductor substrate where a single site in the substrate hosts a magnetic impurity modeled by adding to that site the interaction $\tilde{J} \vect{S} \vect{s}_{\vect{x}_0}$, where $\vect{S}$ is the impurity spin, $\vect{s}_{\vect{x}_0}$ is the local spin of the substrate's electronic system at the impurity site $\vect{x}_0$, and $\tilde{J}$ the coupling strength between the two. We then treat the magnetic impurity in the classical spin approximation, where $\vect{S} \rightarrow \infty$ and $\tilde{J} \rightarrow 0$, such that the product $\tilde{J} \vect{S} \rightarrow \tilde{\vect{J}} $ remains finite.
This approximation is especially suitable for large impurity spins or for systems with a strong anisotropy that aligns the spin in a particular direction \cite{zitko_quantum_2018} and, importantly for our purposes, the model is known to fundamentally capture the influence of a magnetic impurity on the superconductor \cite{salkola1997spectral, morr_impurities_2006, meng2015superconducting,pershoguba2015currents,bjornson2017superconducting}.
As a consequence, the impurity Hamiltonian is now given by 
\begin{equation}
\label{eqn:hamiltonian_imp}
         \tilde{\Hamil}_{i} = \sum_{\sigma, \sigma'} \left(\tilde{\vect{J}} \left(\Vec{\sigma} \right) _{\sigma \sigma'} + \tilde{\varepsilon}_{i} \left(\sigma_0 \right) _{\sigma \sigma'} \right) c^{\dagger}_{\vect{x}_0 \sigma} c_{\vect{x}_0 \sigma'}  ,
\end{equation}
where $c^{\dagger}_{\vect{x}_0 \sigma}$ ($c_{\vect{x}_0 \sigma}$) describes the creation (annihilation) operator of an electron with spin $\sigma$ at the impurity site $\vect{x}_0$ of the substrate.
Without loss of generality, we choose the classical impurity spin $\tilde{\vect{J}}$ to point along the spin-$z$ axis, i.e.~$\tilde{\vect{J}} = \tilde{J} \hat{\vect{z}}$. 
Left as parameters describing the impurity are, therefore, $\tilde{J}$ and $\tilde{\varepsilon}_{i}$.
Using this impurity model in a superconductor with a constant order parameter $\Delta_{\vect{i}} = \Delta_0$ results in the energies of the two in-gap YSR states being \cite{luh1965bound, rusinov1969superconductivity, balatsky_impurity-induced_2006}
\begin{equation}
    \frac{\tilde{\epsilon}_{YSR}}{\Delta_0} = \pm \frac{ (1/\pi N_0)^2 + \tilde{\varepsilon}_{i}^2 - (\tilde{J}/2)^2}{\sqrt{\left[(1/\pi N_0)^2+ \tilde{\varepsilon}_{i}^2 - ( \tilde{J}/2)^2\right]^2 + \tilde{J}^2}},
\end{equation}
where $N_0$ is the normal state density of states at the Fermi level.
The resulting energy spectrum has the same behavior as schematically shown in Fig.~\ref{fig:toy_model}b).
What is left now is to allow $\Delta_{\vect{i}}$ to self-consistently vary in space in response to the magnetic impurity and study the changes of the YSR states.

For the self-consistent calculations, we model the substrate using a tight-binding Hamiltonian on a large two-dimensional (2D) square lattice.
We treat the superconducting interactions using the mean-field approximation for a spin-singlet $s$-wave order parameter within the real-space Bogolioubov-de~Gennes formalism.
The resulting Hamiltonian for the substrate is thus
\begin{equation}
\label{eqn:hamiltonian_sc}
     \tilde{\Hamil}_{sc} = \mu \sum_{\vect{i}, \sigma} c^{\dagger}_{\vect{i} \sigma} c_{\vect{i} \sigma} - t \sum_{\langle\vect{i},\vect{j}\rangle, \sigma} c^{\dagger}_{\vect{i} \sigma} c_{\vect{j} \sigma} +  \sum_{\vect{i}} \Delta_{\vect{i}}  c_{\vect{i} \uparrow} c_{\vect{i} \downarrow} + \text{h.c.},
\end{equation}
where $t$ is the hopping integral between two neighboring sites $\langle \vect{i},\vect{j}\rangle$, which we also use as the unit of energy. 
Further, $\mu$ denotes the onsite energy, which we set to $\mu = 0.5\,t$ to move the Fermi level away from the van Hove singularity at $\mu =0$, in order to model a generic band structure.
We also perform checks for other values of $\mu$, including at the van Hove singularity at $\mu = 0$ and for $\mu \approx 3.62\,t$ where the Fermi level becomes circular.
We find quantitatively similar results, except for a different scaling of the energies related to the order parameter $\Delta_{\vect{i}}$, which can be attributed to a change of the density of states (DOS) at the Fermi level.
This is in line with previous work studying the effects of the van Hove singularity on YSR states, where the YSR energies were found to depend on the change in the zero-energy DOS \cite{uldemolins_effect_2021, basak_shiba_2022}.

Superconductivity is captured by the site-dependent order parameter $\Delta_{\vect{i}}$. We solve fully  self-consistently for $\Delta_{\vect{i}}$ using the self-consistency equation 
\cite{gennes_superconductivity_2019, Black-SchaffePhysRevB.78.024504, bjornson_vortex_2013, reis_self-organized_2014, bjornson_probing_2015, awoga_disorder_2017, bjornson2017superconducting, theiler2019majorana}
\begin{equation}
\label{eqn:gap_equation}
\begin{split}
    \Delta_{\vect{i}} &= - V_{sc}/2 \left( \left \langle c_{\vect{i} \downarrow} c_{\vect{i} \uparrow} \right \rangle - \left \langle c_{\vect{i} \uparrow} c_{\vect{i} \downarrow} \right \rangle \right) \\
    &= - V_{sc}/2 \sum_{\nu} f_T( \epsilon_\nu) \left( v^*_{\nu,\vect{i} \downarrow} u_{\nu, \vect{i} \uparrow} - v^*_{\nu, \vect{i} \uparrow} u_{\nu, \vect{i} \downarrow} \right),
\end{split}
\end{equation}
where $\epsilon_\nu$ denotes the $\nu$-th eigenvalue of the full Hamiltonian $\tilde{\Hamil} = \tilde{\Hamil}_i + \tilde{\Hamil}_{sc}$ and $u_{i\sigma}$, $v_{i\sigma}$ are the corresponding eigenstates at lattice site $i$ for spin $\sigma=\uparrow, \downarrow$. Further, $V_{sc}$ is the effective attractive interaction giving rise to superconductivity, effectively encoding for the attractive (assumed electron-phonon) interaction. We make the reasonable assumption that $V_{sc}$ is constant in the whole sample and not influenced by the single impurity.
The Dirac-Fermi function $f_T(\epsilon)$ generates a temperature $T$ dependence, which we express for convenience in units of $k_B$.

We generate a self-consistent  local order parameter $\Delta_{\vect{i}}$ iteratively, by first diagonalizing $\tilde{\Hamil}$, then calculating \refeqn{eqn:gap_equation}, and finally we use the new $\Delta_{\vect{i}}$ in $\tilde{\Hamil}$ to for next iteration.
As convergence criteria, we accept a solution for $\Delta_{\vect{i}}$ if the smallest difference between two iteration steps for any site is below $5 \cdot 10^{-6}\,\Delta_{0}$, where $\Delta_{0}$ is the order parameter of the unperturbed substrate.
Unless otherwise specified, we employ an attractive potential of $V_{sc} = 1.825\,t$, leading to $\Delta_{0} \approx 0.24\,t$ and a superconducting transition temperature of $T_C \approx 0.14\,t$. For key results we also use other $V_{sc}$ to check the dependence on $V_{sc}$. 
We further limit our results to temperatures well below $T_C$ where the temperature effect on $\Delta_{0}$ itself is negligible.
We use system sizes up to $111 \times 111$ and, find, for the parameters mentioned above, the system to be adequately converged for sizes already of size $61 \times 61$, which is the size we use for our numerical calculations unless otherwise indicated.
To avoid effects from spurious edge states we apply periodic boundary conditions.

To implement the numerical self-consistency calculations we use the Tight-Binding Tool Kit (TBTK) \cite{bjornson_tbtk_2019, TBTK}, using an exact diagonalization solver, which allows us to directly calculate the expectation values necessary for the self-consistency equation \refeqn{eqn:gap_equation}.
We also tested using a kernel polynomial method with Chebychev polynomials \cite{weise_kernel_2006, efficient2010covaci, efficient2012Nagai, bjornson_majorana_2016, mashkoori_impurity_2017, bjornson_tbtk_2019, theiler2019majorana, mashkoori_identification_2020} to speed up the calculations. 
However, this method proved to be unreliable at the very low temperatures we consider, especially for magnetic couplings where the in-gap YSR states are close to zero energy. We  attribute these accuracy problems to the fact that polynomials struggle to capture the sharp Fermi-Dirac function at near-zero temperatures and near zero energy.
While this error can be reduced by increasing the number of coefficients in the polynomial expansion, this diminishes the computational speed-up expected from a kernel polynomial method compared to the, numerically always accurate, exact diagonalization. As a consequence, we instead rely on an exact albeit numerically less efficient diagonalization.


\subsection{Energy spectrum}
\label{subsec:spectrum}
\begin{figure*}
\centering
\includegraphics[width=0.99\textwidth]{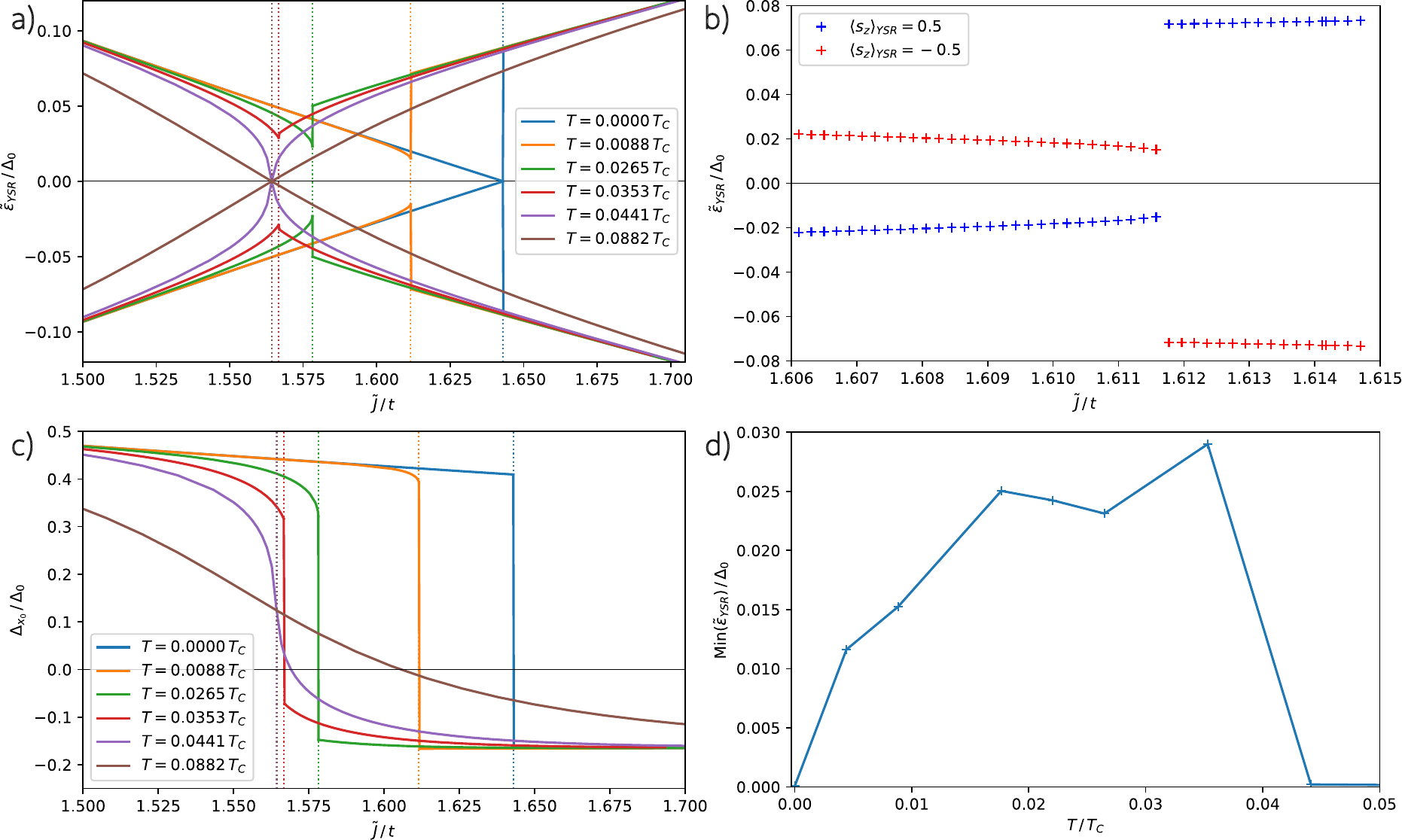}\\%
\caption{
a) YSR state energies $\tilde{\epsilon}_{YSR}$ in a fully self-consistent calculation as a function of coupling $\tilde{J}$ for multiple temperatures. Dotted vertical lines indicate corresponding $\tilde{J}_C$.
b) Zoom-in for single temperature $T = 0.0088\, T_C$, with pluses marking the numerical resolution with respect to $\tilde{J}$. Colors represent the magnetization of each state.
c) Order parameter at the impurity site for increasing $\tilde{J}$ for the same temperatures as in a), with also same dotted vertical lines as in a).
d) Extracted minimum energy of the YSR states (i.e.~energy gap) at $\tilde{J}_C$ as a function of temperature.
}%
\label{fig:SelfconsistentTemperature}%
\end{figure*}

To investigate the influence of calculating the superconducting order parameter self-consistently on the YSR energies, we solve for the energy spectrum around the QPT at $\tilde{J}_C$, using a very dense sampling in $\tilde{J}$ and also varying the temperature. The results are presented in Fig.~\ref{fig:SelfconsistentTemperature}.
In Fig.~\ref{fig:SelfconsistentTemperature}a) we plot the YSR energies as a function of $\tilde{J}$ for different temperatures.
We find a dramatically different behavior for the YSR states compared to the non-self-consistent results in Fig.~\ref{fig:toy_model}b). We further reveal three distinct temperature regimes: zero temperature $T = 0$, a low temperature regime $0 < T < T^*$, and a higher temperature regime $T > T^*$, but where $T^*$ is still much lower than $T_C$.

We start by discussing the zero temperature results in  Fig.~\ref{fig:SelfconsistentTemperature}a).
We find that the YSR states are approximately linearly decreasing (increasing) as a function of $\tilde{J}$ for $\tilde{J}$ below (above) $\tilde{J}_C$. This is fully consistent with the non-self-consistent results around the QPT in Fig.~\ref{fig:toy_model}b).
However, when increasing $\tilde{J}$ and when the states reach zero energy, they discontinuously jump to a finite energy just past the QPT.
Thus, there is a distinct energy jump at $\tilde{J}_C$, albeit, we note technically no gap at zero energy at $T=0$.
We can attribute this energy jump to an equally sudden suppression of the local order parameter $\Delta_{\vect{x}_0}$ at $\tilde{J}_C$, including also a local $\pi$-shift, as shown in Fig.~\ref{fig:SelfconsistentTemperature}c) where we plot the local order parameter as a function of $\tilde{J}$ for the same temperatures.
The discontinuous drop in the local order parameter at the impurity site can in turn be directly attributed to the $\ket{\downarrow}^+$ state suddenly becoming occupied for $\tilde{J} > \tilde{J}_C$ instead of the previously occupied $\ket{\uparrow}^-$ state. 
It has previously been established that the contribution of the $\ket{\downarrow}^+$ to $\Delta_{\vect{x}_0}$ in the self-consistent gap equation Eq.~\refeqn{eqn:gap_equation} is out-of-resonance with the rest of the condensate, or equivalently, $\pi$-shifted compared to other states occupied states \cite{bjornson2017superconducting}.
Hence, suddenly transferring occupation from the $\ket{\uparrow}^-$ state to the $\ket{\downarrow}^+$ state at $\tilde{J}_C$, the local order parameter is equally suddenly suppressed. This sudden suppression is the reason behind the  jump in the YSR energies at $\tilde{J}_C$.
We note that $\pi$-shift in the local order parameter is further energetically favored at the QPT, as it lowers the free energy of the system $F(\tilde{J})$.
We discuss this further in the Appendix \ref{subsec:free_energy}.
We also find no hysteresis behavior at the QPT, as we can self-consistently only stabilize a single order parameter solution for each $J$.
While not evident in Fig. \ref{fig:SelfconsistentTemperature}a), we further observed that $\tilde{J}_C$ is shifted to a lower value compared to the constant $\Delta$ approximation, which is in line with previously reported results \cite{salkola1997spectral, flatte1997localMagnetic, flatte1997localDefects}.
To summarize, the discontinuous jump of the superconducting order parameter and the discontinuous of the YSR energies, both at $\tilde{J}_C$, is intimately connected through the self-consistency gap equation Eq.~\eqref{eqn:gap_equation}.

At small, but finite temperatures we find the same linearly decreasing (increasing) YSR state energies for $\tilde{J}$ far below (above) $\tilde{J}_C$.
In fact, the YSR state energies far below or above $\tilde{J}_C$ are insensitive to temperature. However, close to $\tilde{J}_C$ we find notable temperature effects.
In the temperature regime $0 < T < T^*$, with $T^* \approx 0.044\,T_C$ for the parameters used here, the YSR state energies do not cross through zero energy anymore but always remain at a finite value at $\tilde{J}_C$.
The YSR energy levels still present similarities with the jump present at $T=0$, but the jump is decreasing with increasing temperature such that it is entirely smoothed out for temperatures above $T^*$. When the jump decreases, the gap instead increases for increasing temperatures.
In Fig.~\ref{fig:SelfconsistentTemperature}b) we additionally plot a zoom-in around the QPT to display the finite temperature behavior with each data point corresponding to a separate calculation with different $\tilde{J}$. The figure clearly shows both the gap and the jump at $\tilde{J}_C$ is not an effect of too sparse sampling along the $\tilde{J}$-axis.
Additionally, in color we show the magnetization of the two YSR states $\left\langle s_Z \right\rangle_{YSR}$ to highlight that also the character of the occupied YSR state changes abruptly at $\tilde{J}_C$.
We can understand the behavior of both the jump and the gap by recognizing that finite temperature causes a finite occupation of the $\ket{ \downarrow}^+$ state due to thermal fluctuations even at $\tilde{J}<\tilde{J}_C(T=0)$ (and likewise, not full occupation of the $\ket{ \uparrow}^-$ state). 
As a consequence, the suppression of the local order parameter and its accompanied $\pi$-shift occurs already at lower $\tilde{J}_C$ compared to at $T=0$. This in turn leads to the QPT occurring at a lower $\tilde{J}_C$, i.e.~$\tilde{J}_C(T>0) < \tilde{J}_C(T=0)$, as clearly seen in Fig.~\ref{fig:SelfconsistentTemperature}a).
This move of $\tilde{J}_C$ further needs to be reconciled with the fact that the effects of thermal occupation is only a very low-energy phenomena, and therefore the YSR energies are still at the same energies for all temperatures away from the immediate neighborhood of the QPT. To accommodate both of these behaviors, with minimal other changes to the spectrum,  the energy jump at $\tilde{J}_C$ becomes smaller at finite temperatures, while at the same time a gap necessarily develops at the QPT.
For the parameters chosen in Fig.~\ref{fig:SelfconsistentTemperature}, the energy jump is always to more negative energies for the occupied state when increasing $J$ past  $\tilde{J}_C$ (and inversely for the unoccupied state). 
However, this jump in energy at $\tilde{J}_C$ depends on the trend of the energy spectrum before and after the QPT and therefore, for other parameters this jump in energy can change direction, but is still aways present.
To summarize, the low-temperature behavior is driven by thermal occupation, which decreases $\tilde{J}_C$ with increasing temperature, which in turn forces the spectrum to reduce the energy jump and instead develop an energy gap at $\tilde{J}_C$.

As the effect of thermal fluctuations increases with temperature, we find that the YSR energy gap increases and the jump decreases, along with a decrease in the critical coupling $\tilde{J}_C$.
Finally, as the temperature approaches $T^*$, we find that the discontinuous jump altogether disappears and we are left with only a gap and a kink in the YSR spectrum. For temperatures $T>T^*$, the gap then suddenly closes and there is left only a smooth YSR crossing. This behavior is mimicking that of the local order parameter: At $T^*$ the discontinuous jump in $\Delta_{\vect{x}_0}$ completely disappears and there is then only a smooth evolution of $\Delta_{\vect{x}_0}$ across the QPT. As a consequence, the YSR states can no longer host any drastic changes such as a finite jump, and for temperatures $T \geq T^*$ also the energy gap closes. Overall, the results clearly show that it is the discontinuous behavior of the local order parameter that drives the opening of a gap in the YSR spectrum, and when this is lost due to higher temperatures, the gap also disappears.

Finally, for temperatures much above $T^*$, the YSR state energies again change smoothly through zero energy at $\tilde{J}_C$, similar to the constant $\Delta$ approximation and hence no effect of self-consistency is observed for these higher temperatures. Also, $\tilde{J}_C$ does not change with temperature in this high-temperature regime. Furthermore, the local order parameter $\Delta_{\vect{x}_0}$ does not exhibit a jump at $\tilde{J}_C$, but instead has a constant positive value.
As the local order parameter $\Delta_{\vect{x}_0}$ is positive at $\tilde{J}_C$, this means, in turn, that the $\pi$-shift is moved towards magnetic couplings $\tilde{J}$ larger than $\tilde{J}_C$, as also seen in Fig.~\ref{fig:SelfconsistentTemperature}c). Thus, the existence of the YSR energy gap is not directly related to the $\pi$-shift of the order parameter. The gap is instead tied to the sudden jump in the order parameter, which when it disappears, the YSR gap also disappears, while the $\pi$-shift still remain, albeit its onset is beyond $\tilde{J}_C$.

Additionally, we find that, regardless of temperature, the local suppression of the order parameter does not notably affect the superconducting energy gap in the spectrum, with any change at least a magnitude smaller than the energy gap between the YSR states. Similar effects have also been discussed or shown in \cite{salkola1997spectral, flatte1997localMagnetic, flatte1997localDefects, meng2015superconducting}.

To summarize, the YSR state energies show a complex temperature behavior at $\tilde{J}_C$: a sharp energy jump at $T=0$ develops partially into an energy gap at finite temperatures, until at some temperature scale both jump and gap disappear, leaving a linear zero-energy crossing. All this behavior is intimately tied to the jump in the local superconducting order parameter at $\tilde{J}_C$.
In Fig.~\ref{fig:SelfconsistentTemperature}d) we extract the minimum YSR state energies, i.e.~the YSR state energy gap, as a function of temperature, which shows a notable energy gap for a range of small, but finite temperatures.
To cross-reference our results we also calculate the suppression of the local order parameter in the Anderson impurity model for some key temperatures and find that the $\pi$-shift at the QPT also there causes a finite gap in the YSR spectrum at low temperatures.

\subsection{Magnetization}
\label{subsec:magnetization]}

\begin{figure*}%
\centering
\includegraphics[width=0.95\textwidth]{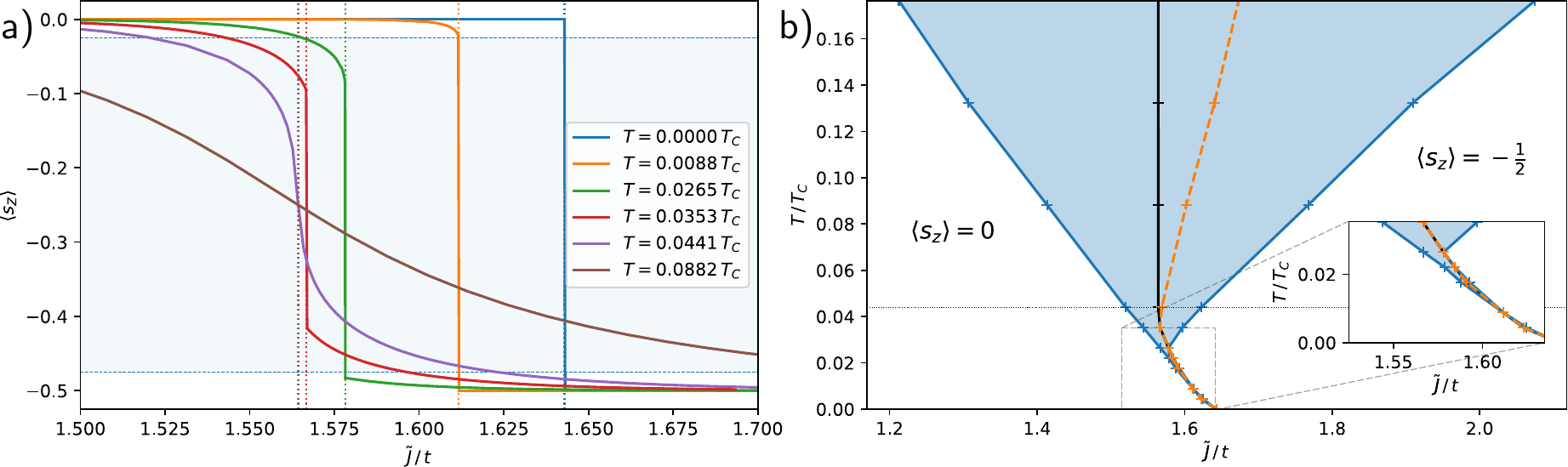}\\%
\caption{ a) Total magnetization of the superconducting ground state $\left\langle s_z \right\rangle$ around $\tilde{J}_C$ for the same temperatures as in \ref{fig:SelfconsistentTemperature}a),c). Vertical dotted lines denote $\tilde{J}_C$. 
b) Phase diagram tracking the change in magnetization for different temperatures as a function of $\tilde{J}$, with the transition region (blue shade) defined as the region with magnetization between $5\%$ and $95\%$ of $\left\langle S_Z \right \rangle = -\frac{1}{2}$.
 Solid black line denotes $\tilde{J}_C$ and dotted orange line denotes $\tilde{J}_\pi$. 
Dotted horizontal line marks the temperature $T^*$. Zoom-in shows behavior at lowest temperatures.}
\label{fig:magnetization_phasediagram}%
\end{figure*}

With the spin orientation of the YSR states being a key component to track their evolution, we can further elucidate the YSR state behavior around the QPT by analyzing the total magnetization of the ground state, which is the expectation value of the operator
\begin{equation}
    \label{eqn:magnetization}
    \vect{s} = \frac{1}{2} \sum_{i, \sigma, \sigma'}  c^\dagger_{i,\sigma} \Vec{\sigma}_{\sigma, \sigma'} c_{i,\sigma'}.
\end{equation}
It is worth noting, that this operator does not include the spin of the impurity, as the impurity is only modeled as an effective Zeeman field acting on one site.
Therefore, the expectation value of $\vect{s}$ yields the response of the superconducting condensate to the magnetic impurity.

For $\tilde{J}<\tilde{J}_C$ the ground state of the whole system is that of a spin-singlet BCS-like state with an unscreened external magnetic moment and thus the total electronic spin of the system $\left\langle \vect{s} \right\rangle = 0$ \cite{balatsky_impurity-induced_2006}.
However, for $\tilde{J} > \tilde{J}_C$ the change of occupation of the YSR state changes, further changing the ground state of the system.
In this regime, the electronic ground state magnetization is instead $\left\langle s_z \right\rangle = - \frac{1}{2}$, with the sign being the opposite of $\vect{J}$.
This ground state can be interpreted as the superconducting condensate partially screening the magnetic moment, resulting in an unpaired electron state which thereby contributes to the spin of the system.

In Fig.~\ref{fig:magnetization_phasediagram}a) we numerically extract the magnetization close to the QPT as a function of $\tilde{J}$ for the same temperatures as in Fig.~\ref{fig:SelfconsistentTemperature}a),c). 
We find that the magnetization follows a trend very similar to that of the local order parameter $\Delta_{\vect{x}_0}$ around $\tilde{J}_C$ in all temperature regimes.
Thus, at zero temperature, the magnetization is a step function from zero to $-\frac{1}{2}$ at $\tilde{J}_C$. For small but finite temperatures, this step is still present but becomes less pronounced with increasing temperatures.
For temperatures $T > T^*$, the discontinuity in magnetization disappears and the magnetization instead takes constant value of $\left\langle s_z \right\rangle = - \frac{1}{4}$ at $\tilde{J}_C$ for all temperatures. In the same way as the local superconducting order parameter is linked to the occupation of the YSR state, so is then the magnetization. At zero temperature, the occupation changes abruptly at $\tilde{J}_C$ and hence we have a full-height step function in the magnetization. With increasing temperature, the occupation close to $\tilde{J}_C$ is not full for either YSR state and hence the step is smaller. 
In particular, there is a ``rounding" for $\tilde{J}<\tilde{J}_C$ since this is where the YSR states are closest to zero energy because of the gap developing on the other side of the phase transition, see Fig.~\ref{fig:SelfconsistentTemperature}a). 
Finally, at higher temperatures, we reach a regime in which both YSR states cross at zero energy and thus are equally occupied, thus explaining $\left\langle s_z \right\rangle = - \frac{1}{4}$.

Based on the results in Fig.~\ref{fig:magnetization_phasediagram}a), we can view the drop in magnetization as a measure for the QPT, and equivalent for $\tilde{J}_C$ (see vertical dotted lines), if the temperature of the system is below $T^*$.
On the other hand, if there is no clear drop in magnetization at the QPT, the temperature in the system is above $T^*$. 
In Fig.~\ref{fig:magnetization_phasediagram}b) we show the transition of the system from magnetization $0$ to $-0.5$ for different temperatures and magnetic couplings, with the blue shaded area marking the transition region where the magnetization takes intermediate values (using a cut-off of $5\%$ from the limiting 0 and $-0.5$ values).
Additionally, we plot the critical coupling $\tilde{J}_C$ at different temperatures (solid black line) and also $\tilde{J}_\pi$, which is the coupling where $\pi$-shift of the local order parameter sets in (dashed orange line).
As discussed above, these latter two curves coincide for low temperatures but start to deviate for $T>T^*$, where $T^*$ is marked with the horizontal dotted line.
By choosing to plot our results as in Fig.~\ref{fig:magnetization_phasediagram}b), we expose a behavior that is seemingly similar to a generic phase diagram for quantum phase transitions \cite{vojta2003quantum} with a wide fan extending from the QPT as the temperature is increased. This similar behavior is intriguing as we only include quantum fluctuations at the mean-field level.
However, unlike a typical quantum phase transition phase diagram, the fan region, or intermediate magnetization region for the self-consistent order parameter solution, starts to expands only above temperature $T^*$, not at absolute zero.
This detail is highlighted by the inset in Fig.~\ref{fig:magnetization_phasediagram}b) and is due to the magnetization changing discontinuously for temperatures below $T^*$.
This is in fact a distinguishing feature of the self-consistent order parameter solution over the constant order parameter approximation, as the magnetization only changes discontinuously for $T=0$ for the latter due to it never developing an energy gap for the YSR states.

\subsection{Parameter dependence of gap}
\label{subsec:parameters}

\begin{figure}%
\centering
\includegraphics[width=0.48\textwidth]{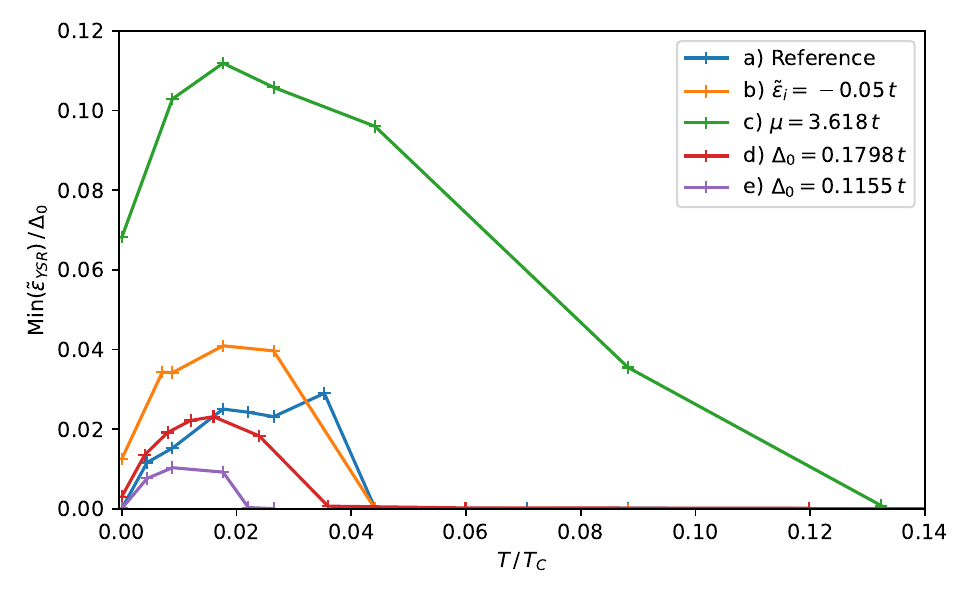}
\caption{
Minimum energy of the YSR states at $\tilde{J}_C$ as a function of temperature, normalized to $T_C$ for different parameters: 
a) Reference case, same as Fig.~\ref{fig:SelfconsistentTemperature}d).
b) Onsite energy added $\tilde{\varepsilon}_{i}=-0.05\,t$.
c) Chemical potential changed to $\mu = 3.618\,t$ but same $\Delta_0$ requiring using $V_{sc} \approx 3.6\,t$.
d) Bulk order parameter decreased to $\Delta_{0} = 0.1798\,t$ using $V_{sc} \approx 1.66\,t$.
e) Bulk order parameter decreased to $\Delta_{0} = 0.1155\,t$ using $V_{sc} \approx 1.48\,t$.
The system sizes for calculations other than a) increased to $81 \times 81$.}
\label{fig:gap_for_different_parameters}%
\end{figure}

Having established the existence of a finite energy gap and also an energy jump for low temperatures, with accompanied behavior in the total magnetization, we finally focus on how the gap behaves with regard to changes in the model parameters. We summarize
the findings in Fig.~\ref{fig:gap_for_different_parameters}, where we plot the gap in the YSR energy spectrum as a function of temperature for a range of parameters. Here we reproduce the result in Fig.~\ref{fig:SelfconsistentTemperature}d) with a blue line to reference the previous discussion.

First of all, for line b) in Fig.~\ref{fig:gap_for_different_parameters} we introduce a small, but finite onsite potential of $\tilde{\varepsilon}_{i} = -0.05\,t$ to the impurity site, which is $10\%$ of the chemical potential $\mu$.
Such local onsite potential can easy be present for an impurity and should generally be included. 
We find that the introduction of this local perturbation in the onsite energy at the impurity site has the overall effect of slightly increased gap for most temperatures.
Importantly, the energies of the YSR states are now also finite at zero temperature. Generally speaking this is an expected development, as we should expect having YSR states touch zero energy before suddenly jumping to finite energies, as in Fig.~\ref{fig:SelfconsistentTemperature}a), to be an unstable situation, energy-wise.
Here, we see that perturbing the impurity with just a slight shift in the onsite potential lifts the zero-energy states and thus generates a gap also at zero temperature. 

We further test our results against changes in the overall chemical potential of the substrate with typical results given by line c) in Fig.~\ref{fig:gap_for_different_parameters}, using $\mu = 3.618\,t$, where the Fermi surface becomes circular in the normal state \cite{graham2017imaging}.
Notably, here we find a substantially increased YSR energy gap, around a factor of three times larger by just changing the Fermi surface.
The range of temperatures in terms of $T_C$, at which the energy gap can be observed at the QPT, is further increased by a similar degree. 
We also observe a finite energy of the YSR states at zero temperature, which we here attribute to an increased particle-hole asymmetry of the normal state, not just at the impurity site.
This can be explained by the fact that particle-hole asymmetry lowers the energy degeneracy of the YSR states at zero energy and thus better facilitates lowering of the total free energy with a gap opening.
In fact, we generally find a stronger suppression of the superconducting order parameter with increased particle-hole asymmetry of the normal state, directly associated with the increased YSR gap.
We, therefore, conclude that considering a self-consistent superconducting order parameter will in general result in an energy gap for the YSR states at all temperatures $T<T^*$, unless a very high degree of particle-hole symmetry is present in the substrate and the impurity. 
We also find that the overall shape of the Fermi surface may influence the size of the gap. This is expected since the Fermi surface is known to influence the properties of the YSR states \cite{flatte1997localDefects, flatte1997localMagnetic, salkola1997spectral, balatsky_impurity-induced_2006, graham2017imaging, basak_shiba_2022}.

Finally, we consider changing the superconducting order parameter value, or equivalently the potential $V_{sc}$.
We here note that calculations with order parameter sizes closer to existing conventional $s$-wave superconductors are hard to achieve numerically, as such small superconducting gaps demand the modeling of very large system sizes, in order to have a reasonable large number of states at the Fermi surface to condense into the superconducting state.
If this is not fulfilled, the results become unstable and inconsistent.
We still try to estimate the effects of smaller superconducting gaps by lowering the order parameters to 75\% and 50\% of the original values for lines d) and e) in Fig.~\ref{fig:SelfconsistentTemperature}d), respectively.
The trends that we extract is that the energy gap is decreased with a decreasing order parameter strengths. We further note that, even for this limited range of order parameters, this trend seems to be non-linear.

In summary, we find that the YSR energy gap at the QPT, and the temperature range over which it can be observed, is strongly dependent on physical parameters non-related to the magnetic moment of the impurity itself, such as the chemical potential, the onsite impurity energy, the Fermi surface shape, and the superconducting bulk order parameter.
Thus, we conclude that the exact value of the energy gap between the YSR states, as well as the transition temperature $T^*$ is highly dependent on the band structure of the substrate and the exact value of the bulk order parameter.
Therefore, our results are qualitative.
For quantitative predictions, the band structure of the material and impurity involved should be taken into account.
In particular, particle-hole asymmetry in the normal state has a tendency to increase the YSR energy gap.

\section{Concluding remarks}
\label{sec:conclusions}

We investigate the YSR state spectrum for a magnetic impurity absorbed on a conventional spin-singlet $s$-wave superconductor, with a focus on the regime around the critical coupling where the system undergoes a quantum phase transition (QPT) from a weakly coupled impurity spin to a screened spin state.
According to tentative experimental observations \cite{karan_superconducting_2022}, 
the YSR states may remain at finite energy at the QPT, an effect that is not predicted by the original YSR theory.
In this work, we identify the only two mechanisms that may cause cause finite energy YSR states at the QPT: by introducing additional interactions to the model description or, by abrupt changes in the system, triggered by the QPT.

We first showed that adding additional interactions on the impurity itself or its bond to the substrate do not cause a finite YSR energy at the QPT.
The only exception is if the substrate lacks inversion symmetry, for example, due to strong disorder, or if the impurity coupling is not fully localized in the substrate and at the same time also breaking inversion symmetry.
In typical experimental configurations, including the one in which finite YSR state energies has been documented \cite{karan_superconducting_2022}, inversion symmetry is seemingly intact and thus additional interactions are not causing the YSR energy gap.
Thus, we conclude that the YSR gap is much more likely generated by the second mechanism, which evokes an abrupt change to the system triggered at the QPT. The only parameter to undergo such an abrupt change is the superconducting order parameter, which self-consistently responds to the YSR states and is thus particularly sensitive to the crossing of YSR states at the QPT. 
Such abrupt changes of the superconducting order parameter at the QPT are already known to exist, often even resulting in a local $\pi$-shift at the impurity site  \cite{salkola1997spectral, flatte1997localDefects, flatte1997localMagnetic, balatsky_impurity-induced_2006, morr_impurities_2006, pershoguba2015currents, graham2017imaging, bjornson2017superconducting, theiler2019majorana}. Our results show that the self-consistently heavily suppressed superconducting order parameter at the impurity also robustly induce a gap in the YSR spectrum below a certain temperature, $T^*$.

To summarize the reason and behavior of the induced gap: at finite temperatures below the temperature $T^*$, there is a discontinuity between the values of the local order parameter for magnetic couplings below and above the critical magnetic coupling at the QPT.
This not only leads to an order parameter with opposite sign at the impurity site, a so-called $\pi$-shift reported previously \cite{salkola1997spectral, flatte1997localDefects, flatte1997localMagnetic, balatsky_impurity-induced_2006, morr_impurities_2006, meng2015superconducting, pershoguba2015currents, graham2017imaging, bjornson2017superconducting, theiler2019majorana}, but also to a shift in the spectrum of the YSR states. This spectral shift causes the YSR states to ``miss" zero energy and, therefore, remain at finite energies even at the critical coupling strength.
Overall, this gives rise to a gap, and often also to a discontinuous jump in the YSR energies.

We further find that the temperature $T^*$ is highly dependent on the model parameters. As some of these parameters are challenging to determine or vary in experimental setups, quantitative predictions about the effects of local suppression of the order parameter and thus the size of the energy gap are difficult to make, given our general model.
Further, given the strong dependence on the chemical potential, we expect that the results also strongly depend on the underlying band structure of the normal state at the Fermi surface.
Still, the qualitative behavior of a strongly temperature-dependent energy gap for the YSR states at the QPT holds for the full range of parameters we explore.

Moreover, the fact that the YSR energy gap generated by the local local order parameter suppression shows a clear dependence on temperature, differentiates it from other possible gap generating terms, in particular, additional putative interactions, as they instead directly change spectral properties of the system, not influenced by temperature.
Therefore, any measured temperature dependence of a YSR  energy gap becomes an indirect, experimentally accessible, sign of the suppression of the local order parameter.

Lastly, we would like to point out that our model description includes fluctuations of the magnetic impurity only at the mean field level. Still, our description yield results that are in good agreement with experimental findings \cite{huang_quantum_2020, karan_superconducting_2022}.
Further including quantum fluctuations of the impurity might modify the results presented in this work.
Therefore, a possible future extension of this work could include those effects, for example via NRG calculations.

\begin{acknowledgments}
We thank J.~Ankerhold, A.~V.~Balatsky, A.~Bouhon, C.~Cuevas, H. Huang, B.~Kubala, and C.~Padurariu for valuable discussions.
AT and AMBS acknowledge financial support from the Knut and Alice Wallenberg Foundation KAW 2019.0309 through the Wallenberg Academy Fellows program.
The computations was enabled by the Berzelius resource provided by the Knut and Alice Wallenberg Foundation at the National Supercomputer Centre (NSC).
All data used to generate the figures in this work is publicly available under \cite{theiler_2025_14825518}.
\end{acknowledgments}

\appendix
\section{Symmetry forbidden gap opening terms}
\label{subsec:YSR_interactions_appendix}
In this Appendix we present results for generating a gap in the YSR state energy spectrum by coupling the two YSR states to each other, when inversion symmetry is not present in the substrate or in the case of a non-inversion symmetric $\vect{k}$ dependent coupling $V_{is}$. This violates the conditions used in the main text for impurity-substrate coupling, and hence these terms are symmetry forbidden in our original model in the main text, but in this Appendix we still consider these possibilities.
Mathematically this means that the summation over $\vect{k}$ in Eq.~\eqref{eqn:coupling_hamiltonian_analytical}, while using the additional terms of $\hat{V}_{is}$~\refeqn{eqn:Vis}, is not causing a full cancellation between individual contributions. 

Below we report the possibilities for all these symmetry forbidden couplings by simply considering each $\vect{k}$ value independently, and thereby we can also suppress the momentum dependence fully in the following. Note though that to reach  the final results we still have to sum over $\vect{k}$, but such sum will also depend on the $\vect{k}$-dependence of $V_{is}$ and can hence not be performed without knowing the details of the interactions.
Further, as discussed in the main text, only coupling terms proportional to $\sigma_1$ and $\sigma_2$ are even potentially capable of opening up a gap in the spectrum.
Also, we find all our results to be identical irrespective of the choice of $\sigma_1$ and $\sigma_2$, and hence, we simplify the notation for the coupling constants to $\alpha_{j} = \alpha_{(1,2)\ts j}$ in the remainder of this Appendix.

We start by considering a coupling that acts on the particle-particle (hole-hole) part of the Nambu space,  i.e.~proportional to $\tau_3$, but now with a spin-dependence in constrast to the $\varepsilon_{i}$ term in Eq.~\eqref{eqn:impurity_level}. Here we find that an additional coupling of the form $\tilde{V}_{is} = \alpha_{3}~\sigma_{1,2} \otimes \tau_3$ produces an effective interaction between the YSR states at the QPT.
To derive analytical results, we keep $\alpha_{3}$ as a generic complex valued scalar and assume that it is much smaller than the superconducting parameter ($\alpha_3 \ll \Delta$) and that the energies of the resulting YSR states are close to zero energy ($E \ll \Delta$). We then find, for an arbitrary linear combination of the $\sigma_1$ and $\sigma_2$ spin orientations, in the interaction the following YSR energies
\begin{widetext}
\begin{equation}
    \epsilon_{YSR}^{\pm, \alpha_{3}} = \pm \Delta \frac{\sqrt{\left( \Gamma^2 + \varepsilon_{i}^2 -J^2\right)^2 + \frac{8 \Re(\alpha_{3})^2 \varepsilon_{i}^2 \Gamma^2}{t_0^2}}}{\sqrt{2 \left(J^2 \Gamma^2 + \Gamma \Delta (2J^2 + 2 \varepsilon_{i}^2 + \Gamma^2) + \Delta^2 (J^2 + \varepsilon_{i}^2 + \Gamma^2) \right)}},
\end{equation}
\end{widetext}
where $\Re$ defines the real part.
From this expression it is straightforward to show that the addition of the term $\alpha_{3}$ generates a gap as soon as an onsite energy $\varepsilon_{i}$ is present, or equivalently, the impurity level has no perfect particle-hole symmetry.
Further assuming $\varepsilon_{i} \ll \Gamma$ and a small order parameter $\Delta \ll \Gamma$, the minimum gap is
\begin{equation} 
    \delta \epsilon_{YSR}^{\alpha_{3}} = \epsilon_{YSR}^{+,\alpha_{3}} - \epsilon_{YSR}^{-,\alpha_{3}} = 4 \Delta \left| \frac{\Re(\alpha_{3}) \varepsilon_{i}}{t_0 \Gamma} \right|.
    \label{eq:gap123}
\end{equation}
This gap appears at the same critical coupling $J_c \approx \Gamma$ as for the unperturbed system.
We find similar dependencies of the minimum YSR energy at the QPT for possible other interactions and summarize the results, both conditions for non-zero gap and the scaling of the gap in Table~\ref{tab:YSR_gaps}.
\begin{table}[ht]
    \caption{Minimal YSR energies at the QPT for different additional couplings $\alpha_{j}$.
    First two columns describe the combinations of spin and Nambu matrices for the additional coupling, third column the physical conditions for the gap to be finite, and the last column the scaling with regards to the interaction strength $\alpha_{j}$.
Results are derived within the following approximation: $\Delta \ll \Gamma$, $\varepsilon_{i} \ll \Gamma$, $\alpha_{j} \ll \Delta$, and for a critical coupling of $J_C \approx \Gamma$.
        }
    \label{tab:YSR_gaps}
\centering
\begin{tabular}{|c|c|c|c|}
    \hline
     $\sigma$ & $\tau$ & conditions  & scaling of $\delta \epsilon_{YSR}^{\alpha_{j}}$\\
     \hline \hline
     \multirow{4}{3em}{$\sigma_{1,2}$ } &  $\tau_0$ & $\Im(\alpha_{0}) \neq 0$ & $\sim \Delta \Im(\alpha_{0})$ \\ \cline{2-4}
     & $\tau_1$ & $\Re(\alpha_{1}) \neq 0$ & $\sim \Delta \Re(\alpha_{1})$ \\ \cline{2-4}
     & $\tau_2$ & $\Re(\alpha_{2})\neq 0 \, \& \, \mu \neq 0$ & $\sim \Delta \Re(\alpha_{2})^\ddagger$ \\ \cline{2-4}
     & $\tau_3$ & $\Re(\alpha_{3}) \neq 0 \, \& \,  \varepsilon_{i} \neq 0$ & $\sim \Delta \Re(\alpha_{3}) \varepsilon_{i} / t_0\Gamma$  \\ \hline
\end{tabular}
\\ \small{$^\ddagger$ determined numerically.}
\end{table}

The first row of Table~\ref{tab:YSR_gaps} describes a coupling that is spin selective, i.e.~different hopping strengths depending on the spin orientation in the $x,y$-plane.
The second and third rows describe an unconventional superconducting pairing interaction mediated via the impurity-substrate bond.
In the latter case, a gap only opens if the particle-hole symmetry is not present in the substrate, i.e.~the substrates chemical potential $\mu \neq 0$.
For the third row, we have to also determine the scaling of the minimal energy gap numerically, as the analytically extracted local Green's function for the substrate superconductor \refeqn{eqn:local_bare_sc_greens_fct} does not describe this case.
Finally, the fourth row restates the results of Eq.~\eqref{eq:gap123}, which is a spin orbit coupling related.
The results in Table \ref{tab:YSR_gaps} show that a gap in the spectrum always scales linearly with the real or imaginary part of the coupling $\alpha_j$ and the magnitude of the order parameter $\Delta$. Finally, we again emphasize that all these gaps are for a single $\vect{k}$, and as soon as inversion symmetry is present in the substrate, they will all still cancel to zero. It is only when such inversion symmetry is not present that the gaps in Table \ref{tab:YSR_gaps} may materialize.

\section{Free energy across QPT}
\label{subsec:free_energy}
\begin{figure}%
\centering
\includegraphics[width=0.475\textwidth]{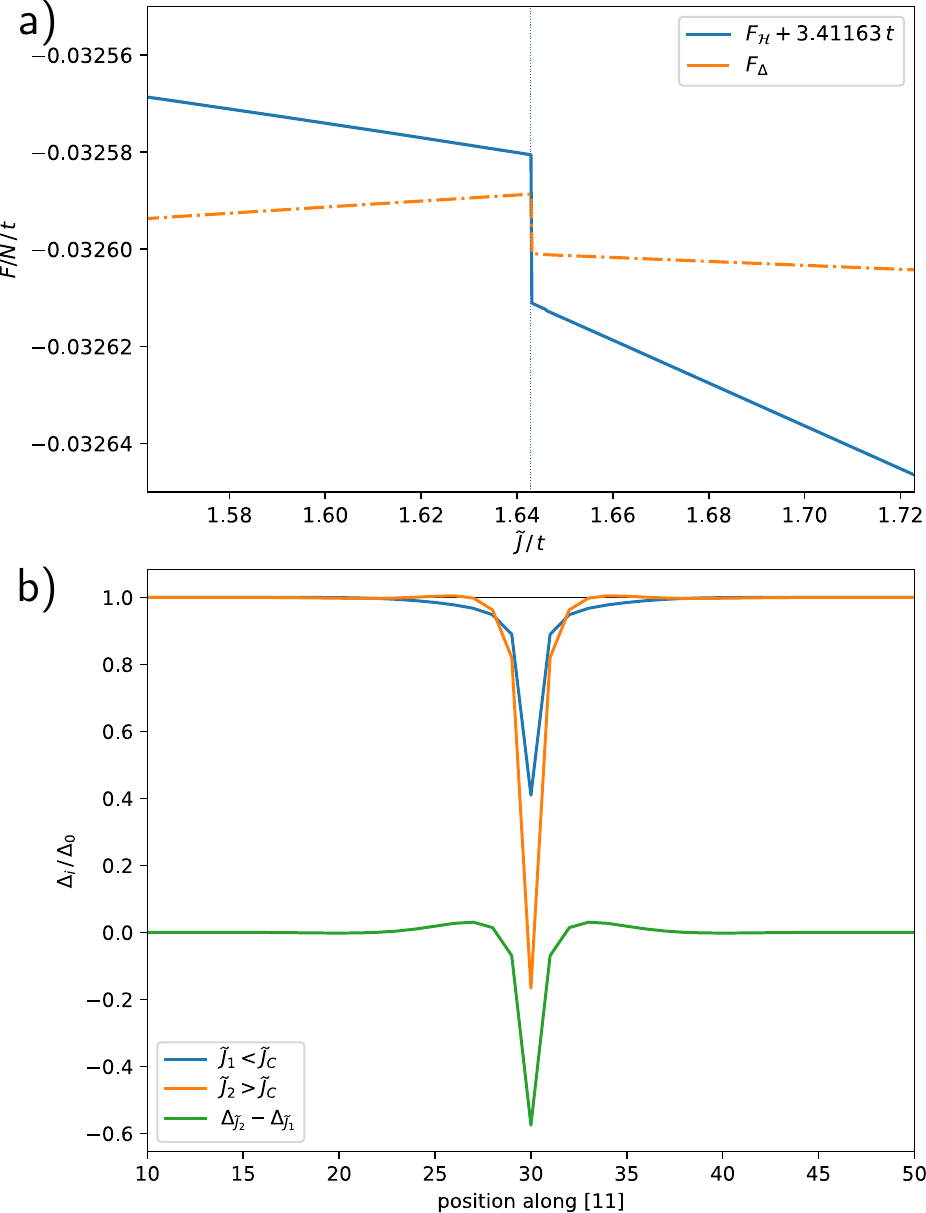}\\%
\caption{
a) Free energy contributions $F_{\tilde{\Hamil}}$ and $F_{\Delta}$ at zero temperature as a function of $\tilde{J}$ and normalized by the number of sites $N$. 
Dashed vertical line denotes $\tilde{J}_C$.
b) Spatial behavior of the local order parameter along the [11] direction. Two values, $\tilde{J}_1$ and $\tilde{J}_2$, are chosen to be just below and above $\tilde{J}_C$ within the numerical resolution of $\tilde{J}$. Solid black line indicate bulk $\Delta$ (for $\tilde{J}=0)$.
Green line shows the difference in $\Delta_{\vect{i}}$ between these two cases.}
\label{fig:FreeEnergy}%
\end{figure}

In this Appendix, we provide a complement to the main discussion about energy spectrum and magnetization by also elucidating the behavior of the total free energy around the QPT. At zero temperature, the free energy of the system is the sum of the two contributions $F = F_{\tilde{\Hamil}} + F_{\Delta}$:
\begin{equation}
\label{eqn:free_energy_components}
    \begin{split}
        F_{\tilde{\Hamil}} &= \sum_{i} f_{0}(\epsilon_i) \epsilon_i \\
        F_{\Delta} &= -\frac{1}{V_{Sc}} \sum_{\vect{i}} \left|\Delta_{\vect{i}} \right|^2,
   \end{split}
\end{equation}
where the first contribution $F_{\tilde{\Hamil}}$ comes from the energy spectrum of the Hamiltonian $\tilde{\Hamil}$ given the sum of the Hamiltonians in  Eq.~\refeqn{eqn:hamiltonian_sc} and Eq.~\refeqn{eqn:hamiltonian_imp}, and with $f_0$ being the zero temperature Dirac-Fermi function. 
The second part, $F_{\Delta}$, accounts for additional energy contribution of the superconducting order parameter due to its mean-field decomposition.
For both of these contributions, we find a jump at the QPT  at $\tilde{J}_C$, significantly lowering the energy of the ground state for $\tilde{J}>\tilde{J}_C$, see Fig. \ref{fig:FreeEnergy}a).
The drop in $F_{\tilde{\Hamil}}$ is largest and is a direct consequence of jump of the YSR energies at $\tilde{J}_C$. The drop in $F_\Delta$ is smaller, where a drop, instead of a raise, may at first seem  surprising since $\Delta$ is drastically suppressed at the impurity site.
To understand this we plot in \ref{fig:FreeEnergy}b) the local order parameter extracted along the [11] direction, cutting through the impurity.
There is a strong local suppression of the order parameter together with a $\pi$-shift at the impurity site, which itself yields an increase in the free energy for $\tilde{J} > \tilde{J}_C$ compared to $\tilde{J} < \tilde{J}_C$. However, we find that the suppression of the local order parameter becomes more localized and the order parameter is even slightly enhanced in the vicinity of the impurity for magnetic coupling is above $\tilde{J}_C$. Similar ``ripples" have been reported before for magnetic impurities \cite{schlottmann1976spatial, salkola1997spectral,flatte1997localDefects, flatte1997localMagnetic}.
Their exact location is due to the YSR state localization being dependent on $v_F$.
Overall, we find that these effects actually outweigh the local suppression of $\Delta_{\vect{x}_0}$. It is most pronounced along the [11] direction of the lattice but exists along other directions as well.
This directional dependence is due to the normal-state Fermi surface.
As expected, for a fully circular Fermi surface, we find a circular suppression/enhancement of $\Delta_\vect{i}$.
We observed a similar behavior for the free energy for finite temperatures, while the additional free energy contribution of the entropy is negligible.

%

\end{document}